\documentclass[doublecol]{epl2}
\usepackage{amsfonts,amsmath,amssymb,graphicx,bm}

\title{Possible singlet and triplet superconductivity on honeycomb lattice}
\shorttitle{Possible singlet and triplet superconductivity on honeycomb lattice}

\author{Long-Yun Xiao\inst{1,2} \and Shun-Li Yu\inst{1,2} \and Wei Wang\inst{1,2} \and Zi-Jian Yao\inst{3} \and Jian-Xin Li\inst{1,2}}
\shortauthor{L.-Y. Xiao \etal}

\institute{
  \inst{1} National Laboratory of Solid State Microstructures and Department of Physics, Nanjing University, Nanjing 210093, China \\
  \inst{2} Collaborative Innovation Center of Advanced Microstructures, Nanjing University, Nanjing 210093, China \\
  \inst{3} Department of Physics and Institute of Theoretical Physics, Nanjing Normal University, Nanjing, 210023, China
}

\pacs{74.20.Rp}{Pairing symmetries (other than s-wave)}
\pacs{74.20.Mn}{Nonconventional mechanisms}
\pacs{71.10.Fd}{Lattice fermion models (Hubbard model, etc.)}

\abstract{We study the possible superconducting pairing symmetry mediated by
spin and charge fluctuations on the honeycomb lattice using the
extended Hubbard model and the random-phase-approximation method. From $2\%$ to $20\%$ doping levels, a spin-singlet $d_{x^{2}-y^{2}}+id_{xy}$-wave is shown to be the leading superconducting pairing symmetry when only the on-site Coulomb interaction $U$ is considered, with the gap function being a mixture of the nearest-neighbor and next-nearest-neighbor pairings. When the offset of the energy level between the two sublattices exceeds a critical value, the most favorable pairing is a spin-triplet $f$-wave which is mainly composed of the next-nearest-neighbor pairing. We show that the next-nearest-neighbor Coulomb interaction $V$ is also in favor of the spin-triplet $f$-wave pairing.}

\begin{document}

\maketitle

Since the production of graphene (a honeycomb lattice of carbon atoms) in 2004 \cite{KSNovoselov}, the realization of superconductivity on the honeycomb lattice have attracted considerable interest \cite{SIchinokura,JChapman,BMLudbrook,JLMcChesney}. Recently, the studies on the Ca-intercalated bilayer graphene and the graphene laminates observed superconductivities at 4 K \cite{SIchinokura} and 6.4 K \cite{JChapman} respectively. Furthermore, another recent experimental study also presented evidence for superconductivity in Li-decorated monolayer graphene with the transition temperature around $5.9$ K \cite{BMLudbrook}. On the theoretical side, the studies have been extended to models of interacting electrons on the honeycomb lattice, without necessarily concentrating on the parameter regions relevant to graphene, as other systems based on this geometry have been found \cite{AMBlack-Schaffer}. Especially, nitrides $\beta$-MNCl (M=Hf,Zr) which are composed of alternate stacking of honeycomb layers have been observed to exhibit superconductivity with $T_{c}\sim15$ K for Zr \cite{SYamanaka} and $T_{c}\sim25$ K for Hf \cite{SYamanaka1} by doping carriers. Various experimental results, including a weak isotope effect \cite{HTou,YTaguchi} and the $T$-linear specific heat \cite{YTaguchi1}, have pointed to an unconventional superconducting state, and the magnetic susceptibility measurements \cite{YKasahara} suggest that the electron pairings are possibly  mediated by magnetic fluctuations in these materials.

Many theoretical studies based on the Hubbard model predict a superconducting order parameter with $d_{x^{2}-y^{2}}+id_{xy}$ symmetry in the spin-singlet channel at half filling and low doping levels \cite{BUchoa,CHonerkamp,SPathak,TMa,MLKiesel,RNandkishore,RNandkishore1,WWu,Black-Schaffer,XYXu}, while a recent study with the variational cluster approximation and the cellular dynamical mean field theory suggests that the dominant pairing is a spin-triplet with the $p_{x}+ip_{y}$ symmetry \cite{JPLFaye}. A variational-Monte-Carlo (VMC) study shows that both the $d_{x^{2}-y^{2}}$-wave and $d_{x^{2}-y^{2}}+id_{xy}$-wave are the possible superconducting pairing symmetry, but the state with $d_{x^{2}-y^{2}}$-wave symmetry has the larger condensation energy \cite{TWatanabe}. Another quantum-Monte-Carlo study predicts that the favored state would have $p_{x}+ip_{y}$ symmetry in the spin-triplet channel but at a large doping level ($\sim80\%$) \cite{TMa1}. Overall, the pairing symmetry of the possible superconductivity of the interacting electron system on the honeycomb lattice is still under debate. In this paper, motivated by the experimental and theoretical progresses, we investigate the superconducting parings mediated by spin and charge fluctuations on the honeycomb lattice by using the extended Hubbard model [including both the on-site interaction $U$ and nearest-neighbor (NN) interaction $V$] and the random-phase-approximation (RPA) method. The doping concentration is set from $2\%$ to $20\%$. We find that the spin-singlet $d_{x^{2}-y^{2}}+id_{xy}$-wave is the leading superconducting pairing when $V=0$. Our results reveal that the electron pairing in the singlet channel is composed of both the NN and next-nearest-neighbor (NNN) pairings. When the off-site Coulomb interaction $V$ exceeds a critical value, we find that the dominant pairing is a spin-triplet $f$-wave mainly composed of the NNN pairings. We then study the effect of the offset of the energy level between the two sublattice and find that the energy-level offset is in favor of the spin-triplet pairing. Our results also indicates that the triplet pairing is more easily realized in the low doping range.

\begin{figure}
  \centering
  \includegraphics[scale=0.5]{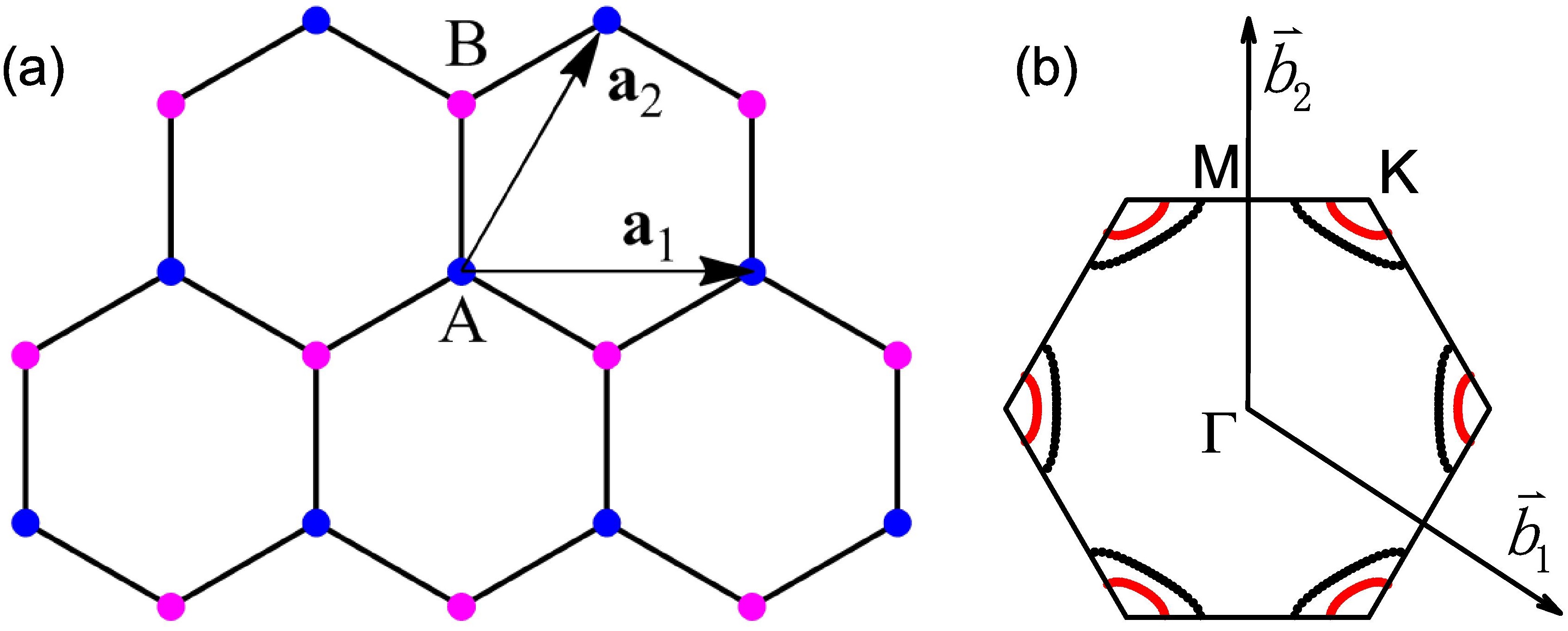}
  \caption{(color online) (a) Structure of honeycomb lattice. $A$ and $B$ denote the two sublattices, $\mathbf{a}_{1}$ and $\mathbf{a}_{2}$ are the translation vectors. (b) FSs in the first BZ for the $5\%$ (the red lines) and $15\%$ (the black lines) dopings at $\Delta_{s}=0$. $\mathbf{b}_{1}$ and $\mathbf{b}_{2}$ are the reciprocal-lattice vectors.}
  \label{fig1}
\end{figure}

The structure of honeycomb lattice is shown in Fig. \ref{fig1}. There are two inequivalent lattice sites labeled by $A$ and $B$ respectively. The model Hamiltonian contains two parts:
\begin{equation}
H=H_{0}+H_{int}.
\end{equation}
The bare Hamiltonian $H_{0}$ reads
\begin{equation}
H_{0}=-t\sum_{\langle ij\rangle\sigma}(a_{i\sigma}^{\dag}b_{j\sigma}+h.c.)+\Delta_{s}(\sum_{i\sigma}n^{a}_{i\sigma}-\sum_{j\sigma}n^{b}_{j\sigma}),
\label{bare-h}
\end{equation}
where $a_{i\sigma}$ ($a_{i\sigma}^{\dag}$) annihilates (creates)
an electron with spin $\sigma$ ($\sigma=\uparrow,\downarrow$) on
site $\bm{r}_{i}$ of the sublattice $A$ (an equivalent definition
is used for the sublattice B) and
$n^{a}_{i\sigma}=a_{i\sigma}^{\dag}a_{i\sigma}$
($n^{b}_{j\sigma}=b_{j\sigma}^{\dag}b_{j\sigma}$),
$\langle\cdot\cdot\rangle$ denotes the NN bond.
Since a single-layer honeycomb lattice in $\beta$-MNCl consisting
of alternating ``M" and ``N" sites \cite{SYamanaka,SYamanaka1}, which induces an offset of the
energy level between the two sublattices, we include the
$\Delta_{s}$ terms in $H_{0}$ to study the effects of the
energy-level offset. The interactions between electrons in
$H_{int}$ include the on-site and NN Coulomb interactions, i.e.
\begin{equation}
H_{int}=U(\sum_{i}n^{a}_{i\uparrow}n^{a}_{i\downarrow}+\sum_{j}n^{b}_{j\uparrow}n^{b}_{j\downarrow})+V\sum_{\langle ij\rangle}n^{a}_{i}n^{b}_{j},
\end{equation}
where $n^{a}_{i}=n^{a}_{i\uparrow}+n^{a}_{i\downarrow}$ ($n^{b}_{j}=n^{b}_{j\uparrow}+n^{b}_{j\downarrow}$).

Based on the scenario that the pairing interaction arises from the exchange of spin and charge fluctuations, we can calculate the effective electron-electron interaction using the RPA. The spin-singlet pairing interaction is given by \cite{TakimotoT}
\begin{equation}
\hat{V}^{s}(q)=\frac{3}{2}\hat{U}^{s}\hat{\chi}^{s}(q)\hat{U}^{s}-\frac{1}{2}\hat{U}^{c}\hat{\chi}^{c}(q)\hat{U}^{c}+\frac{1}{2}(\hat{U}^{c}+\hat{U}^{s}),
\label{singlet-v}
\end{equation}
while the spin-triplet pairing interaction is given by \cite{TakimotoT}
\begin{equation}
\hat{V}^{t}(q)=-\frac{1}{2}\hat{U}^{s}\hat{\chi}^{s}(q)\hat{U}^{s}-\frac{1}{2}\hat{U}^{c}\hat{\chi}^{c}(q)\hat{U}^{c},
\label{triplet-v}
\end{equation}
where $\hat{\chi}^{s}$ ($\hat{\chi}^{c}$) is the spin (charge) susceptibility and $\hat{U}^{s}$ ($\hat{U}^{c}$) is the interaction matrix for the spin (charge) fluctuation. $\hat{\chi}^{s}$ and $\hat{\chi}^{c}$ are expressed as $\hat{\chi}^{s}(q)=[1-\hat{\chi}^{0}(q)\hat{U}^{s}]^{-1}\hat{\chi}^{0}(q)$ and $\hat{\chi}^{c}(q)=[1+\hat{\chi}^{0}(q)\hat{U}^{c}]^{-1}\hat{\chi}^{0}(q)$ respectively. The non-interacting susceptibility is given by $\hat{\chi}^{0}_{\mu\nu,\eta\varphi}(q)=-\frac{T}{N}\sum_{k}G_{\eta\mu}(k+q)G_{\nu\varphi}(k)$ with the number of lattice sites $N$ and temperature $T$. Here, $\hat{V}^{s/c}$, $\hat{\chi}^{s/c}$ and $\hat{\chi}^{0}$ are $4\times4$ matrices. $\mu$, $\nu$, $\eta$ and $\varphi$ are the sublattice indices. The matrix multiplications in Eqs. (\ref{singlet-v}) and (\ref{triplet-v}) are defined as $(\hat{A}\hat{B})_{\mu\nu,\eta\varphi}=\sum_{\alpha\beta} A_{\mu\nu,\alpha\beta}B_{\alpha\beta,\eta\varphi}$. The Green's function is a $2\times2$ matrix and given by $\hat{G}(k)=[\mathrm{i}\omega_{n}-\hat{H}_{0}(\bm{k})+\mu]^{-1}$. In the above, $k\equiv(\bm{k},i\omega_{n})$ with $\omega_{n}=(2n+1)\pi T$, $q\equiv(\bm{q},i\omega_{m})$ with $\omega_{m}=2n\pi T$. By performing the Matsubara frequency summation, the susceptibility can be written as
\begin{align}
\hat{\chi}^{0}_{\mu\nu,\eta\varphi}(\bm{q},i\omega_{m})=&\frac{1}{N}\sum_{\bm{k},ij}\frac{u_{i}^{\eta}(\bm{k})u_{i}^{\mu\ast}(\bm{k})u_{j}^{\nu}(\bm{k}+\bm{q})u_{j}^{\varphi\ast}
(\bm{k}+\bm{q})}{i\omega_{m}-E_{i}(\bm{k})+E_{j}(\bm{k}+\bm{q})} \nonumber \\
&\times[f(E_{i}(\bm{k}))-f(E_{j}(\bm{k}+\bm{q}))].
\end{align}
where $u^{\mu}_{i}(\bm{k})=\langle\mu,\bm{k}|i,\bm{k}\rangle$ projects the band basis $|i,\bm{k}\rangle$ to the sublattice basis $|\mu,\bm{k}\rangle$. Here, $i$ and $\mu$ are the band and sublattice index respectively. $E_{i}(\bm{k})$ is the energy of the Hamiltonian (\ref{bare-h}) for the band $j$ at the momentum $\bm{k}$, and $f(E)$ is the Fermi distribution function.
$\hat{U}^{s}$ and $\hat{U}^{c}$ are given by: for $\mu=\nu=\eta=\varphi$,
$U^{s}_{\mu\nu,\eta\varphi}=U$ and $U^{c}_{\mu\nu,\eta\varphi}=U$; for $\mu=\nu=1$ and $\eta=\varphi=2$, $U^{s}_{\mu\nu,\eta\varphi}=0$ and $U^{c}_{\mu\nu,\eta\varphi}=Ve^{ik_{y}}+2V\cos(\frac{\sqrt{3}}{2}k_{x})e^{-i\frac{1}{2}k_{y}}$; for $\mu=\nu=2$ and $\eta=\varphi=1$, $U^{s}_{\mu\nu,\eta\varphi}=0$ and $U^{c}_{\mu\nu,\eta\varphi}=Ve^{-ik_{y}}+2V\cos(\frac{\sqrt{3}}{2}k_{x})e^{i\frac{1}{2}k_{y}}$; for other cases, $U^{s}_{\mu\nu,\eta\varphi}=0$ and $U^{c}_{\mu\nu,\eta\varphi}=0$.

The linearized superconducting gap equation (the ``Eliashberg" equation) is given by \cite{TakimotoT}
\begin{align}
\lambda\Delta_{mn}(k)&=-\frac{T}{N}\sum_{q}\sum_{\eta\varphi}\sum_{\mu\nu}
V^{s/t}_{\eta m,n\varphi}(q)G_{\eta\mu}(k-q) \nonumber\\
&\times G_{\varphi\nu}(q-k)\Delta_{\mu\nu}(k-q).
\label{eliashberg-eq}
\end{align}
We confine our considerations to the dominant scattering occurring in the vicinity of the FS. Thus, we can reduce the effective interactions (\ref{singlet-v}) and (\ref{triplet-v}) together with the ``Eliashberg" equation (\ref{eliashberg-eq}) to the Fermi surface (FS). The scattering amplitude of a Cooper pair from the state $(\bm{k},-\bm{k})$ on the FS of band $i$ to the state $(\bm{k}^{\prime},-\bm{k}^{\prime})$ on the FS of band $j$ is calculated from the projected interaction \cite{SGraser,Shunli-1}
\begin{align}
\Gamma_{ij}(\bm{k},\bm{k}^{\prime})\!=\!\!\sum_{\mu\nu\eta\varphi}&u^{\mu\ast}_{i}(-\bm{k})u^{\nu\ast}_{i}(\bm{k})V^{s/t}_{\varphi\nu,\mu\eta}(\bm{k}-\bm{k}^{\prime},\omega\!=\!0) \nonumber \\
&\times u^{\eta}_{j}(\bm{k}^{\prime})u^{\varphi}_{j}(-\bm{k}^{\prime}),
\end{align}
We then solve the following eigenvalue problem \cite{SGraser,Shunli-1}:
\begin{equation}
-\sum_{j}\oint_{C_{j}}\frac{d\bm{k}^{\prime}_{\|}}{4\pi^{2}|\nabla E_{j}(\bm{k}^{\prime})|}\Gamma_{ij}(\bm{k},\bm{k}^{\prime})g_{j}(\bm{k}^{\prime})=\lambda g_{i}(\bm{k}),
\label{eigeq}
\end{equation}
where $g_{i}(\bm{k})$ is the normalized gap function along the FS of band $i$. The integral in Eq. (\ref{eigeq}) is evaluated along the FSs. The most favorable SC pairing symmetry corresponds to the gap function with the largest eigenvalue $\lambda$. One merit of this method is that it can adequately include the effect of DOS on the FS \cite{Shunli-1}.

\begin{figure}
  \centering
  \includegraphics[scale=0.65]{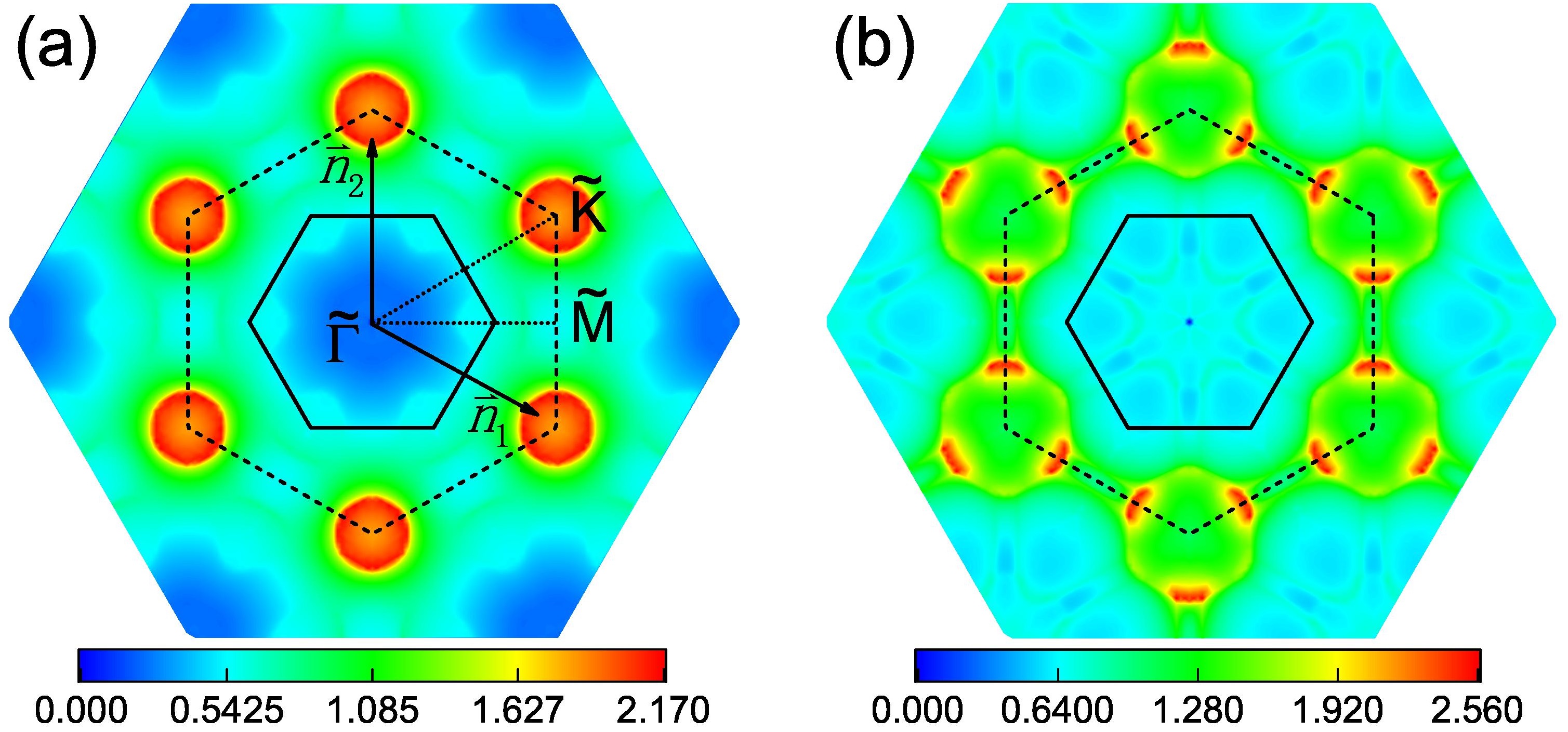}
  \caption{(color online) Spin susceptibilities under $U=2.0t$, $V=0$ and $\Delta_{s}=0$  for $5\%$ (a) and $15\%$ doping (b) respectively. The solid lines indicate the first BZ, while the dashed lines indicate the minimum repeating unit of spin susceptibility.}
  \label{fig2}
\end{figure}

To solve the eigenequation (\ref{eigeq}), we use $128$ points along each FS. The temperature is set at $T=0.005$, and the calculation of the susceptibility is done with uniform $128\times128$ meshes. Since at half filling (one electron per site), the Fermi level is located at the Dirac point [$K$ point in Fig. \ref{fig1}(b)], for which the DOS vanishes, we introduce carriers into this system to make the superconductivity more easily realized. As the Hamiltonian has the particle-hole symmetry, we only study the hole-doping case.  The fermiology for the two typical dopings is presented in Fig. \ref{fig1}(b).

\begin{figure}
  \centering
  \includegraphics[scale=0.7]{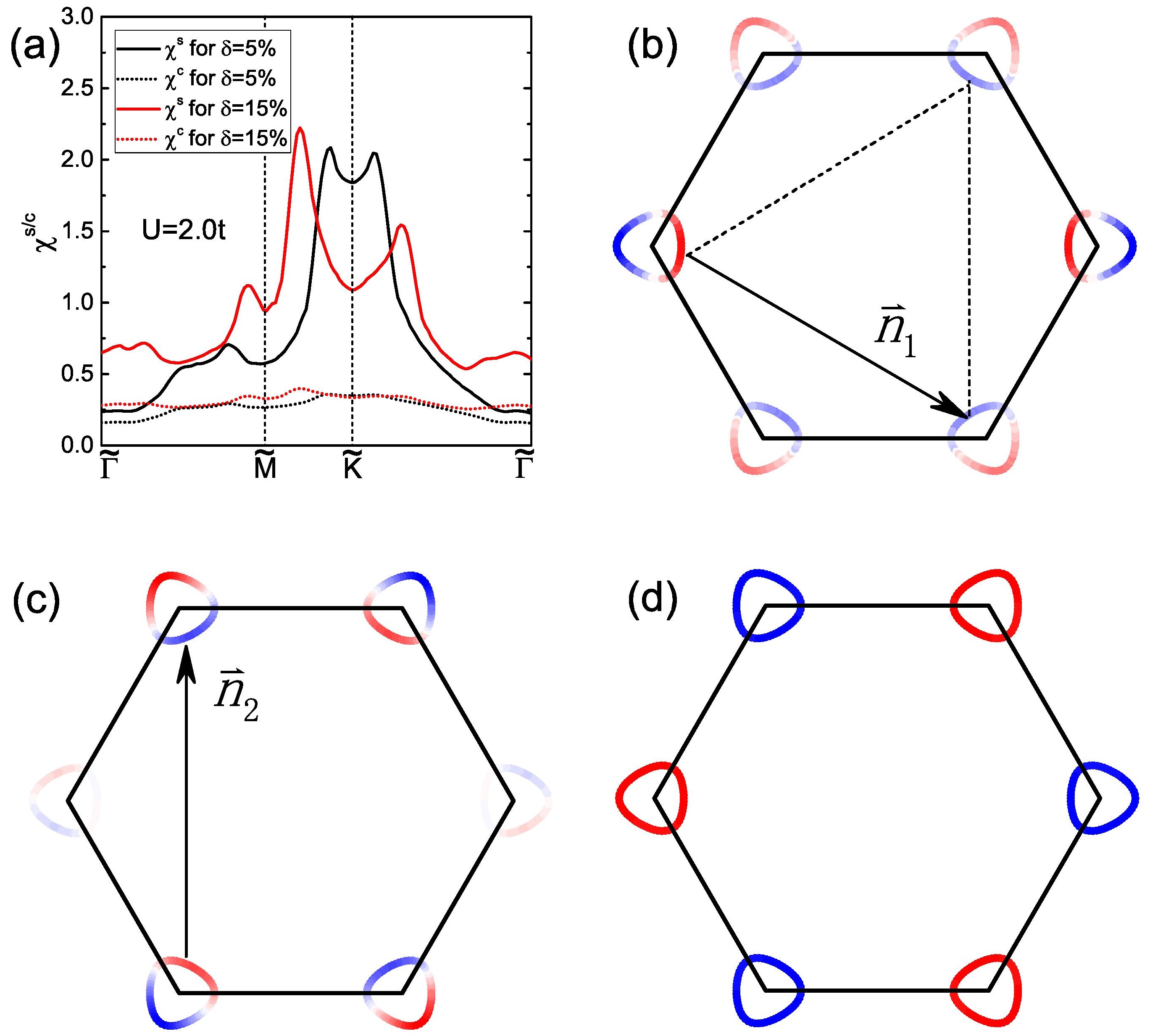}
  \caption{ (color online) (a) Spin and charge susceptibilities along the symmetric lines for $5\%$ and $15\%$ doping. The points $\tilde{\Gamma}$, $\tilde{M}$ and $\tilde{K}$ are indicated in Fig. \ref{fig2}(a). (b) and (c) are the dominant pairing functions in the spin-singlet channel for $\delta=5\%$. (d) Dominant pairing function in the spin-triplet channel for $\delta=5\%$. Here, $U=2.0$, $V=0$ and $\Delta_{s}=0$.}
  \label{fig3}
\end{figure}

The pairing interactions (\ref{singlet-v}) and (\ref{triplet-v}) are directly determined by the spin and charge susceptibilities $\hat{\chi}^{s}$ and $\hat{\chi}^{c}$. In Fig. \ref{fig2}(a), we present the static spin fluctuations $\tilde{\chi}^{s}(\bm{q})=\sum_{\mu\nu}\chi^{s}_{\mu\mu,\nu\nu}(\bm{q},\omega=0)$ for $5\%$ doping with $U=2.0$ and $V=0$. It is worth noting that as shown in Fig. \ref{fig2}(a), the spin susceptibility $\tilde{\chi}^{s}(\bm{q})$ is not periodic with period $\bm{b}_1$ and $\bm{b}_2$, which are the reciprocal primitive vectors of the honeycomb lattice as shown in Fig. \ref{fig1}(b). The two-sublattice structure introduces a phase difference upon a translation of the reciprocal primitive vector, which results in a larger periodic unit cell of the inter-sublattice spin susceptibilities\cite{StauberT,Stauber}. We can find that the peaks form a ring structure as indicated by the vectors $\vec{n}_{1}$ and $\vec{n}_{2}$. At a higher doping level such as $15\%$ doping, the ring structures are changed into some patch structures [see Fig. \ref{fig2}(b)].
If we continue to increase the doping concentration to $25\%$, the Fermi level will be at the singularity point and the FS will have perfect nesting property, which will result in the chiral spin-density-wave instability \cite{TLi,WSWang,SJiang,MakogonD}. This type of chiral spin-density wave has also been proposed on triangular and kagome lattices \cite{IMartin,SLYu1}. As our purpose is to study the superconductivity, we choose the doping in the range from $2\%$ to $20\%$.
As for the charge susceptibility, our results show that it has a similar
structure as the spin susceptibility but is much less in magnitude [see Fig. \ref{fig3}(a)]. Thus, when $V=0$, we concentrate on the effect of spin fluctuations in discussing the pairing symmetry.

\begin{figure}
  \centering
  \includegraphics[scale=0.6]{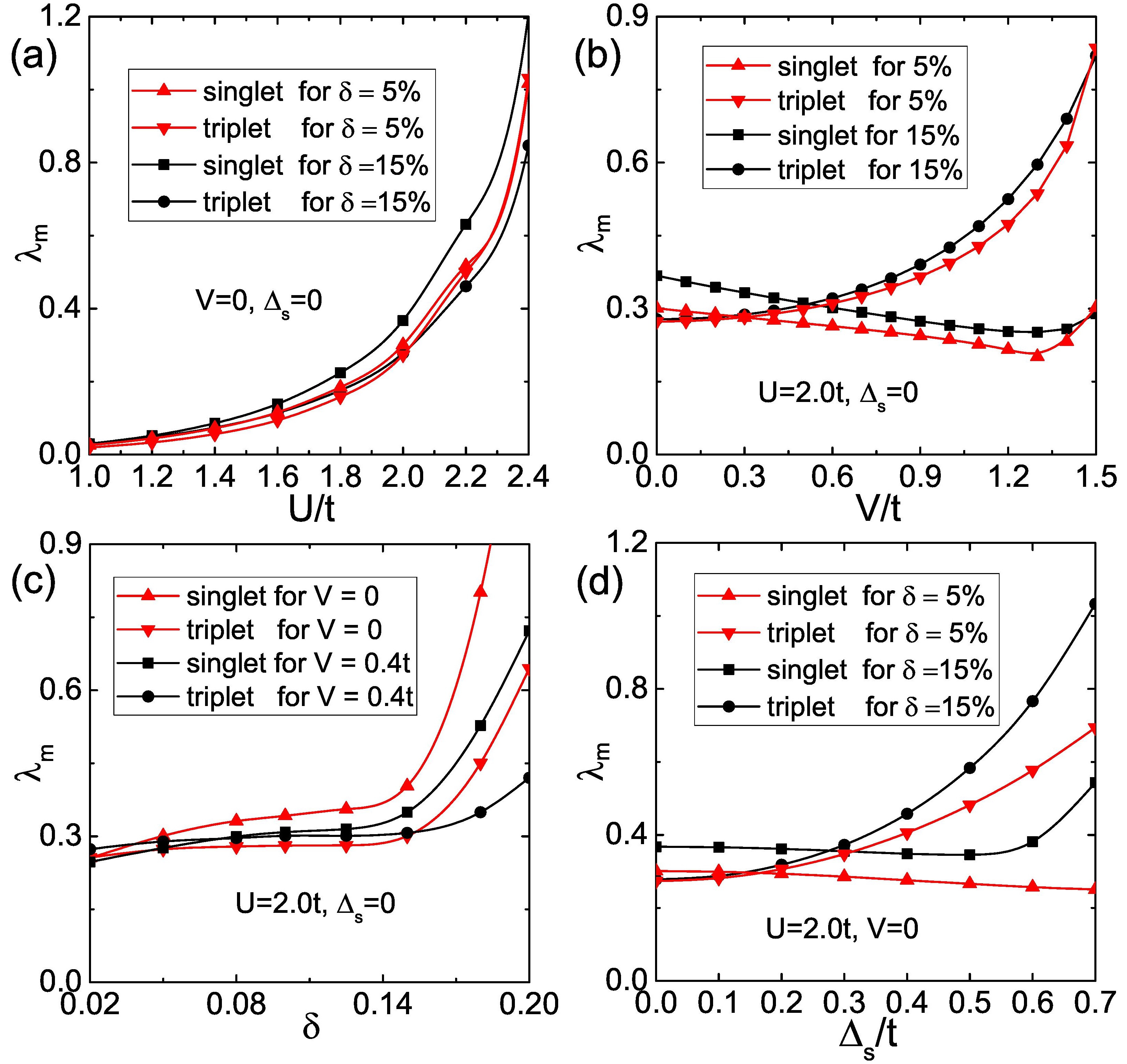}
  \caption{ (color online) (a) $U$ dependence of the maximum eigenvalue $\lambda_{m}$ of the gap equation at $V=0$ and $\Delta_{s}=0$ for $\delta=5\%$ and $\delta=15\%$. (b) $V$ dependence of $\lambda_{m}$ at $U=2.0t$ and $\Delta_{s}=0$. (c) Doping dependence of $\lambda_{m}$ at $U=2.0t$ and $\Delta_{s}=0$. (d) $\Delta_{s}$ dependence of $\lambda_{m}$ at $U=2.0t$ and $V=0$.}
   \label{fig4}
\end{figure}

The most favorable pairing symmetry can be obtained by finding out
$\bm{g}(\bm{k})$ with the largest eigenvalue $\lambda_{m}$
from Eq. (\ref{eigeq}). When $V=0$ and $\Delta_{s}=0$, we find that the dominant pairing is in
the spin-singlet channel and is two-fold degenerate. These two orthogonalized pairing functions are presented in Fig. \ref{fig3}(b) and (c), where one is $d_{x^{2}-y^{2}}$-like and the other is $d_{xy}$-like. We can infer that the system will realize the $d_{x^{2}-y^{2}}+id_{xy}$ superconducting state,
which gives a fully gapped qusiparticle spectrum \cite{Black-Schaffer}. We remark that the obtained gap functions deviate a lot from the ``standard" forms (which is usually assumed to be exclusively the NN or NNN pairings) that have been assumed in many previous literatures \cite{BUchoa,CHonerkamp,SPathak,TMa,MLKiesel,RNandkishore,RNandkishore1,WWu,Black-Schaffer,XYXu}. In the scenario of the ``Eliashberg" equation, the order parameter is obtained self-consistently without presuming its structure in real-space or momentum-space. This deviation indicates that the actual realized superconducting gap function could be more complex than the ``standard" form. We will address this issue later in the context of real-space pairing.
In analogy with the analysis that has been applied to the discussions on
the $d$-wave pairing in high-$T_{c}$ cuprates \cite{DJScalapino}
and the $s_{\pm}$-wave pairing in iron-based superconductors
\cite{ZJYao,SLYu}, we can see that if the spin susceptibility
has a peak around a special wave-vector $\bm{Q}$, the pair
scatterings from $(\bm{k},-\bm{k})$ to
$(\bm{k}+\bm{Q},-\bm{k}-\bm{Q})$, which is the channel for Cooper pairing, will be dominating the scattering process. Because the pairing interaction for
the spin-singlet pairing $\hat{V}^{s}(\bm{Q})$ [Eq.
(\ref{singlet-v})] is positive, the pairing function will try to satisfy
the condition $g_{i}(\bm{k})g_{j}(\bm{k}+\bm{Q})<0$ to
ensure the largest eigenvalue $\lambda$ of the Eq. (\ref{eigeq}). In other words, the pairing functions connected by the vectors $\vec{n}_{1}$ and $\vec{n}_{2}$ are expected to have opposite signs [see Fig. \ref{fig3}(b) and (c)]. However, for the present system, there are conflicts in satisfying the condition $g_{i}(\bm{k})g_{j}(\bm{k}+\bm{Q})<0$ [indicated by the
triangle in Fig. \ref{fig3}(b)] due to the special fermiology. Actually, this is the reason that why the pairing instabilities in spin-singlet and spin-triplet channels are closely competing with each other [see Fig. \ref{fig4}(a)]. We note that for $15\%$
doping, the structures of gap functions are qualitatively consistent with those for $5\%$ doping (not shown here).

Though the honeycomb lattice has two energy bands, the electron pairing mainly comes from the intra-band pairing in doping systems,
the inter-band pairing is significantly smaller as it is not on-shell energywise. The transformation of gap function between the sublattice and band representations is given by
\begin{align}
\Delta_{ij}(\bm{k})=\sum_{\eta\varphi}u^{\eta}_{i}(\bm{k})u^{\varphi}_{j}(-\bm{k})\Delta_{\eta\varphi}(\bm{k}),
\label{gap_transformation}
\end{align}
where $i$ and $j$ are the band indices, $\mu$ and $\varphi$ are the sublattice indices. From Eq. (\ref{gap_transformation}), we can see that the pair function on the band comes from both intra-sublattice and inter-sublattice pairings. The dominant intra-sublattice and inter-sublattice pairings respectively come from the NN and NNN pairing in the real space [see Fig. \ref{fig1}(a)]. Considering that the pairing intensities are the same along the bonds belonging to the same type (NN or NNN) but
with phase differences for different directions, we will get the
following pairing functions in the spin-singlet channel:
\begin{align}
\Delta^{(1)}_{ab}(\bm{k})&=e^{-ik_{y}}+e^{i\theta}e^{i(\frac{\sqrt{3}}{2}k_{x}+\frac{1}{2}k_{y})} \nonumber \\
&+e^{i\theta^{\prime}}e^{-i(\frac{\sqrt{3}}{2}k_{x}-\frac{1}{2}k_{y})}
\end{align}
for the NN pairing which satisfies $\Delta^{(1)}_{ab}(\bm{k})=\Delta^{(1)}_{ba}(-\bm{k})$, and
\begin{align}
\Delta^{(2)}_{aa}(\bm{k})&=\cos(\sqrt{3}k_{x})+e^{i\theta}\cos(\frac{\sqrt{3}}{2}k_{x}+\frac{3}{2}k_{y}) \nonumber \\
&+e^{i\theta^{\prime}}\cos(\frac{\sqrt{3}}{2}k_{x}-\frac{3}{2}k_{y})
\end{align}
for the NNN pairing which satisfies $\Delta^{(2)}_{aa}(\bm{k})=\Delta^{(2)}_{aa}(-\bm{k})$ and $\Delta^{(2)}_{aa}(\bm{k})=\Delta^{(2)}_{bb}(\bm{k})$. Here, $a$ and $b$ denote the $A$ and $B$ sublattices, and we set the length of NN $AB$ bond as $1$. The most natural choices for the phases are $(\theta,\theta^{\prime})=(0,0)$ and $(\theta,\theta^{\prime})=\pm(2\pi/3,4\pi/3)$, which are corresponding to the $s$ and $d+id$ symmetries respectively. In the following, we set $(\theta,\theta^{\prime})=(2\pi/3,4\pi/3)$ and perform a linear combination of $\Delta^{(1)}$ and $\Delta^{(2)}$ to fit the numerical results discussed above. The fitting function is
\begin{align}
\Delta_{11}(\bm{k})&=\alpha^{(1)}[u^{a}_{1}(\bm{k})u^{b}_{1}(-\bm{k})\Delta^{(1)}_{ab}(\bm{k}) \nonumber \\
&+u^{a}_{1}(-\bm{k})u^{b}_{1}(\bm{k})\Delta^{(1)}_{ab}(-\bm{k})] \nonumber \\
&+2\alpha^{(2)}[u^{a}_{1}(\bm{k})u^{a}_{1}(-\bm{k})\Delta^{(2)}_{aa}(\bm{k}) \nonumber \\
&+u^{b}_{1}(\bm{k})u^{b}_{1}(-\bm{k})\Delta^{(2)}_{bb}(\bm{k})],
\end{align}
where $\Delta_{11}(\bm{k})$ is the pairing function of the lower band as we consider the case of hole doping, $\alpha^{(1)}$ and $\alpha^{(2)}$ are the variational parameters to fit the numerical results. We then project $\Delta_{11}(\bm{k})$ on the FS and adjust $\alpha^{(1)}$ and $\alpha^{(2)}$ to get the maximum overlap between $\Delta_{11}(\bm{k})$ and the calculated results. We find the overlap can be up to $95\%$ when $|\alpha^{(1)}/\alpha^{(2)}|=1.3:1$. This ratio implies that the spin-singlet pairing is a mixture of both NN and NNN pairings in real space. Our results suggest that the acutally realized superconducting gap function may well deviate from the ``standard" form of $d_{x^{2}-y^{2}}+id_{xy}$.

In the same spirit, we discuss the real-space pairings in the spin-triplet channel. The dominant pairing function is presented in Fig. \ref{fig3}(d), and we find that it has an $f$-wave symmetry. The NN pairing in the triplet channel is
\begin{align}
\Delta^{(1)}_{ab}(\bm{k})&=e^{-ik_{y}}+e^{i\theta}e^{i(\frac{\sqrt{3}}{2}k_{x}+\frac{1}{2}k_{y})} \nonumber \\
&+e^{i\theta^{\prime}}e^{-i(\frac{\sqrt{3}}{2}k_{x}-\frac{1}{2}k_{y})}
\end{align}
which satisfies $\Delta^{(1)}_{ab}(\bm{k})=-\Delta^{(1)}_{ba}(-\bm{k})$, and the NNN pairing is
\begin{align}
\Delta^{(2)}_{aa}(\bm{k})&=\sin(\sqrt{3}k_{x})+e^{i\theta}\sin(\frac{\sqrt{3}}{2}k_{x}+\frac{3}{2}k_{y}) \nonumber \\
&+e^{i\theta^{\prime}}\sin(\frac{\sqrt{3}}{2}k_{x}-\frac{3}{2}k_{y})
\end{align}
which satisfies $\Delta^{(2)}_{aa}(\bm{k})=-\Delta^{(2)}_{aa}(-\bm{k})$ and $\Delta^{(2)}_{aa}(\bm{k})=\Delta^{(2)}_{bb}(\bm{k})$. The natural choices for the phases $(\theta,\theta^{\prime})=(0,0)$ or $(\theta,\theta^{\prime})=\pm(2\pi/3,4\pi/3)$ are corresponding to the $f$ and $p+ip$ symmetries respectively. According to the numerical results, we set $(\theta,\theta^{\prime})=(0,0)$. Then, the fitting function is
\begin{align}
\Delta_{11}(\bm{k})&=\alpha^{(1)}[u^{a}_{1}(\bm{k})u^{b}_{1}(-\bm{k})\Delta^{(1)}_{ab}(\bm{k}) \nonumber \\
&-u^{a}_{1}(-\bm{k})u^{b}_{1}(\bm{k})\Delta^{(1)}_{ab}(-\bm{k})] \nonumber \\
&+2\alpha^{(2)}[u^{a}_{1}(\bm{k})u^{a}_{1}(-\bm{k})\Delta^{(2)}_{aa}(\bm{k}) \nonumber \\
&+u^{b}_{1}(\bm{k})u^{b}_{1}(-\bm{k})\Delta^{(2)}_{bb}(\bm{k})],
\end{align}
Optimizing $\alpha^{(1)}$ and $\alpha^{(2)}$ to get the maximum overlap between the fitting function and the numerical results, we find that the NNN pairing is the leading component, while the NN component is significantly small. Thus, the dominant triplet pairing comes almost entirely from the NNN pairing.

We note that the recent researches have found that the effective on-site interaction $U$ can vary in the range from $1.2t$ to $2.3t$ for graphene, silicene and benzene \cite{WehlingTO,MSchuler}, so we next investigate the evolutions of the superconducting states with the magnitude of interactions. The evolutions of the maximum eigenvalue $\lambda_{m}$ with the Hubbard $U$ for $V=0$ in the spin-singlet and spin-triplet channels for two typical doping concentrations ($\delta=5\%$ and $\delta=15\%$) are presented in Fig. \ref{fig4}(a). We find that the spin-singlet and spin-triplet pairing channels are closely competing with each other at $5\%$ doping, though the singlet pairing channel is favoured slightly. Increasing the doping
level to $15\%$, the singlet and triplet pairing channels becomes well separated. We then examine the effect of the NN interaction $V$ on the superconducting state. Figure \ref{fig4}(b) shows the evolutions of $\lambda_{m}$ with $V$ at $U=2.0t$. We find that the triplet pairing becomes the dominant superconducting instability at $V\approx0.3t$ for $\delta=5\%$ and $V\approx0.5t$ for $\delta=15\%$ respectively, and the favorable pairing state in the triplet channel has an $f$-wave symmetry as shown in Fig. \ref{fig3}(d), which is consistent with the results based on the perturbative functional-renormalization-group calculation \cite{CHonerkamp}. This can be understood as following: from Eqs. (\ref{singlet-v}) and (\ref{triplet-v}), we can see that the effective pairing interaction $\hat{V}^{s}$ for the spin-singlet channel is suppressed with the increase of $V$ (as $\hat{U}^{c}$ is increased consequently), while $\hat{V}^{t}$ for the spin-triplet channel is enhanced, so the spin-triplet pairing is more favorable than the spin-singlet pairing as $V$ is larger than the critical value. On the other hand, we have seen that both the NN and NNN pairings occur in the singlet channel while only the NNN pairing in the triplet channel, so the NN interaction $V$ can suppress the pairing in the singlet channel.

At last, we consider the effects of doping level and energy-level offset $\Delta_{s}$ between the two sublattices. The doping dependence of $\lambda_{m}$ is presented in Fig. \ref{fig4}(c) for $U=2.0t$ and two typical values of $V$ ($V=0$ and $V=0.4t$). We find that the increase of doping concentration is more beneficial to the spin-singlet pairing, and the spin-triplet pairing state is realized more likely in the low doping region for sufficiently large values of $V$ [also see Fig. \ref{fig4}(b)]. This is because the singlet and triplet pairings are nearly degenerate for low dopings [Fig. \ref{fig4}(a)]. From Fig. \ref{fig4}(d) illustrating the $\Delta_{s}$ dependence of $\lambda_{m}$, the energy-level offset $\Delta_{s}$ is shown to favor the triplet pairing. In the case of $\Delta_{s}=0$, the energy bands have equal weight of the two sublattices, but for $\Delta_{s}\neq0$ the distributions of the sublattices are not equal in every band. We have discussed above that the spin-singlet and spin-triplet channels are respectively dominated by the inter-sublattice and intra-sublattice pairings, so the energy-level offset $\Delta_{s}$ will suppress the singlet pairing and enhance the triplet pairing.

In summary, we investigate the superconducting parings mediated by spin and charge fluctuations on the honeycomb lattice by using a spin-fluctuation scenario. The doping concentration is set from $2\%$ to $20\%$. We find that the spin-singlet $d_{x^{2}-y^{2}}+id_{xy}$-wave is the leading superconducting pairing when $V=0$. Our results reveal that the electron pairing in the spin-singlet channel is dominated by both the NN and NNN pairings. The spin-triplet pairing is mainly composed of the NNN pairings. At low dopings and large $V$, we find that the dominant pairing is in the spin-triplet channel and the $f$-wave paring is most favorable. We also study the effect of the offset of the energy level between the two sublattice and find that the energy-level offset is in favor of the spin-triplet pairing.

\acknowledgments
This work was supported by the National Natural Science Foundation of China (11374138, 11190023, 11204125 and 11404163) and Natural Science Foundation of Jiangsu Province (BK20140589).

\end{document}